**From Simulation to Suture Lines: Computational Fluid Dynamic Simulation-Guided Baffle Construction for Complex Biventricular Repair**

**Authors:** Alexander K. Reed, MD[1,2]; Elena Sabdy Martinez[2,3]; Alexander D. Kaiser, PhD[1,2,4]; Amit Sharir[1]; Caio Carcaioli Bonin[1]; Perry S. Choi, MD[1,2,4]; Shiraz A. Maskatia, MD[5]; Alison L. Marsden, PhD[2-6]; and Michael R. Ma, MD[1,2,4]

[1]Department of Cardiothoracic Surgery, Stanford University School of Medicine, Stanford, CA

[2]Cardiovascular Institute, Stanford University, Stanford, CA

[3]Institute for Computational and Mathematical Engineering, Stanford University, Stanford, CA

[4]Maternal & Child Health Research Institute, Stanford University, Stanford, CA

[5]Division of Pediatric Cardiology, Department of Pediatrics, Stanford University School of Medicine, Stanford, CA

[6]Department of Bioengineering, Stanford University, Stanford, CA

**Word count:** 1,000

**Disclosures:** None.

IRB approval through Stanford University IRB #74566 (approved 12/10/25). Patient consent obtained from legal representative.

AKR supported in part by NIH R38HL143615. ADK supported in part by NIH K25HL175208, AHA 24CDA1272816 and Stanford Maternal & Child Health Research Institute. PSC supported in part by the Maternal and Child Health Research Institute as the Stephen Bechtel Endowed Fellow in Pediatric Translational Medicine. ALM and MRM supported in part by NIH R01HL173845.

**Corresponding:**

Michael R. Ma, MD

870 Quarry Rd Extension

Palo Alto, CA 94304

(650) 723-5771

mma@stanford.edu

## Introduction

Biventricular repair is favored for children with complex double-outlet right ventricle (DORV) to eliminate the long-term morbidity of single ventricle physiology. However, creating a baffle that maintains adequate cross-sectional area while limiting incursion on the tricuspid valve apparatus is challenging in a distorted, arrested heart. Three-dimensional (3D) models have aided baffle design with anatomic visualization but analytic approaches, including computational fluid dynamic (CFD) simulation support, have not yet been incorporated into clinical workflows. We present the preoperative, prospective construction of a DORV repair baffle guided by a semi-automated area-preserving algorithm and CFD analysis.

## Methods

A 15-month-old boy with DORV with a subpulmonic ventricular septal defect (VSD) (**Figure 1A**) was referred for biventricular repair. At birth, he had a severely dysplastic pulmonic valve and left pulmonary artery (PA) stenosis and underwent balloon atrial septostomy, left pulmonary artery homograft patch augmentation, main PA ligation, and a 4mm modified Blalock-Taussig-Thomas shunt placement at one month of age. He progressed well until he developed increasing cyanotic episodes. Preoperative cardiac computed tomography angiography (CTA) was performed in accordance with standard clinical care. Images were acquired using dose-modulated retrospective ECG gating. Informed consent was obtained for CFD simulation (Stanford IRB #74566, approved 12/10/25). The end-systolic phase was segmented to generate a three-dimensional computational model for baffle planning and CFD analysis (**Figure 1B**).[1] Segmentation accuracy was confirmed with a pediatric cardiovascular imaging expert. Once complete, the model was reviewed preoperatively with the surgeon to define the margins of the baffle interactively. Using the surgeon-specified suture line and our described methodology[1], a patient-specific baffle template was generated that maintained a minimum target cross-sectional area of 80.12 $mm^2$ which was derived from the mean aortic annular diameter for body surface area (10.1mm, Z-score=0, Boston Z-score database). Critically, this produced a simulation-ready geometry that could be meshed for CFD analysis. For CFD analysis, a healthy reference waveform was scaled to the patient's BSA to prescribe pulsatile flow with a normative mean cardiac index of 4.5 $L/min/m^2$.[2] Pressure differences were measured between the VSD and aortic-annular planes.[1] A 1.25 mm rim was added to account for the portion consumed by the suture line intraoperatively. The baffle was then extruded into a 3D-template, printed with biocompatible resin, and sterilized. We embossed the location of the VSD and the right ventricular surface to ensure correct orientation during implantation. The patient was brought to the operating room, where the surgeon again reviewed the 3D model. Postoperatively, baffle position relative to the planned template was assessed. The implanted baffle was segmented from a postoperative CTA. The pre- and postoperative geometries were aligned using the VSD and aortic cap centers and normals. A least-squares best-fit plane was constructed from the aligned planned baffle boundary, and both boundaries were projected onto this shared plane. Template-to-implant distance was measured from the planned outline to the postoperative baffle boundary.

## Results

The predicted pressure gradient across the computationally planned baffle was 4.39 mmHg (**Figure 1C**). Preoperative computational baffle design greatly facilitated intraoperative baffle construction. Prior to cannulation for cardiopulmonary bypass, the sterile baffle template was brought to the field, and a piece of bovine pericardium was prepared and trimmed exactly to the baffle template (**Figure 1D**). The trimmed pericardium was labeled before being submerged in a saline bath until use.

Following shunt ligation, PA reconstruction, and main PA oversew, a right ventriculotomy was performed and bulging conal septal tissue between the great roots was resected alongside a small portion of interventricular muscular septum. The pre-cut baffle was then implanted with two running 5-0 polypropylene sutures starting at the inferior margin of the VSD (**Figure 1E**). Cross-clamp and cardiopulmonary bypass times were 161 and 306 minutes, respectively, with no reapplication of the cross-clamp and no intraoperative patch revision. The chest was left open to facilitate recovery. The chest was closed on postoperative day (POD) 3; the patient was extubated on POD5. All inotropes were weaned by POD6. Transthoracic echocardiogram on POD18 revealed normal biventricular function, a small intramural VSD at the aortic outlet, and a peak baffle gradient of 2 mmHg, in line with predicted postoperative values. Postoperative CTA demonstrated no significant baffle leak, no encroachment onto the tricuspid valve, and strong concordance of the baffle to the CFD-guided template. Baffle position was measured to be 1.05±0.89 mm from the preoperatively planned model. Baffle perimeter was 99.41 mm, a 1.96% reduction from the preoperative template (**Figure 1F**). The patient was discharged home on POD23.

## Discussion

The use of high-fidelity digital reconstructions of patient anatomy has potential to revolutionize surgical planning in complex congenital heart disease. Here, we apply CFD analysis to preoperative baffle design for biventricular septation with outflow tract reconstruction. While CFD modeling has been applied prospectively for procedural planning to optimize flow in Fontan palliation[3] and aortic reconstruction[4], it has not previously been utilized for intracardiac baffle design. To our knowledge, this is the first report of prospective CFD to generate a template that directly guides intraoperative baffle construction. While we adopted a similar approach to interactive baffle boundary selection to published methods[5], key modifications here allow for semi-automated, area-preserving baffle geometry creation with a CFD simulation-ready output. Preoperative CFD ensures predicted postoperative left ventricular outflow tract performance and pressure gradients are clinically acceptable in the new subaortic anatomy. Use of an intraoperative template also enables implantation of a patch optimized for systole as opposed to traditional patch fabrication intraoperatively when the heart is arrested and flaccid. Furthermore, template use simplifies conceptualization of the baffle on the arrested heart and permits its construction off-pump, thereby theoretically reducing cross-clamp duration, a known predictor of post-cardiotomy shock.[6] Here, we demonstrate strong concordance between our planned template and the implanted baffle. Future simulations should account for anisotropic elasticity of the pericardial patch, as well as cardiac contraction and baffle dynamics throughout the cardiac cycle. In future studies, preoperative CFD will be critical in guiding biomechanically optimized VSD enlargement, conal septal resection, and repositioning of the tricuspid subvalvar apparatus. Future investigation will further quantify potential intra- and postoperative benefits with this technique.

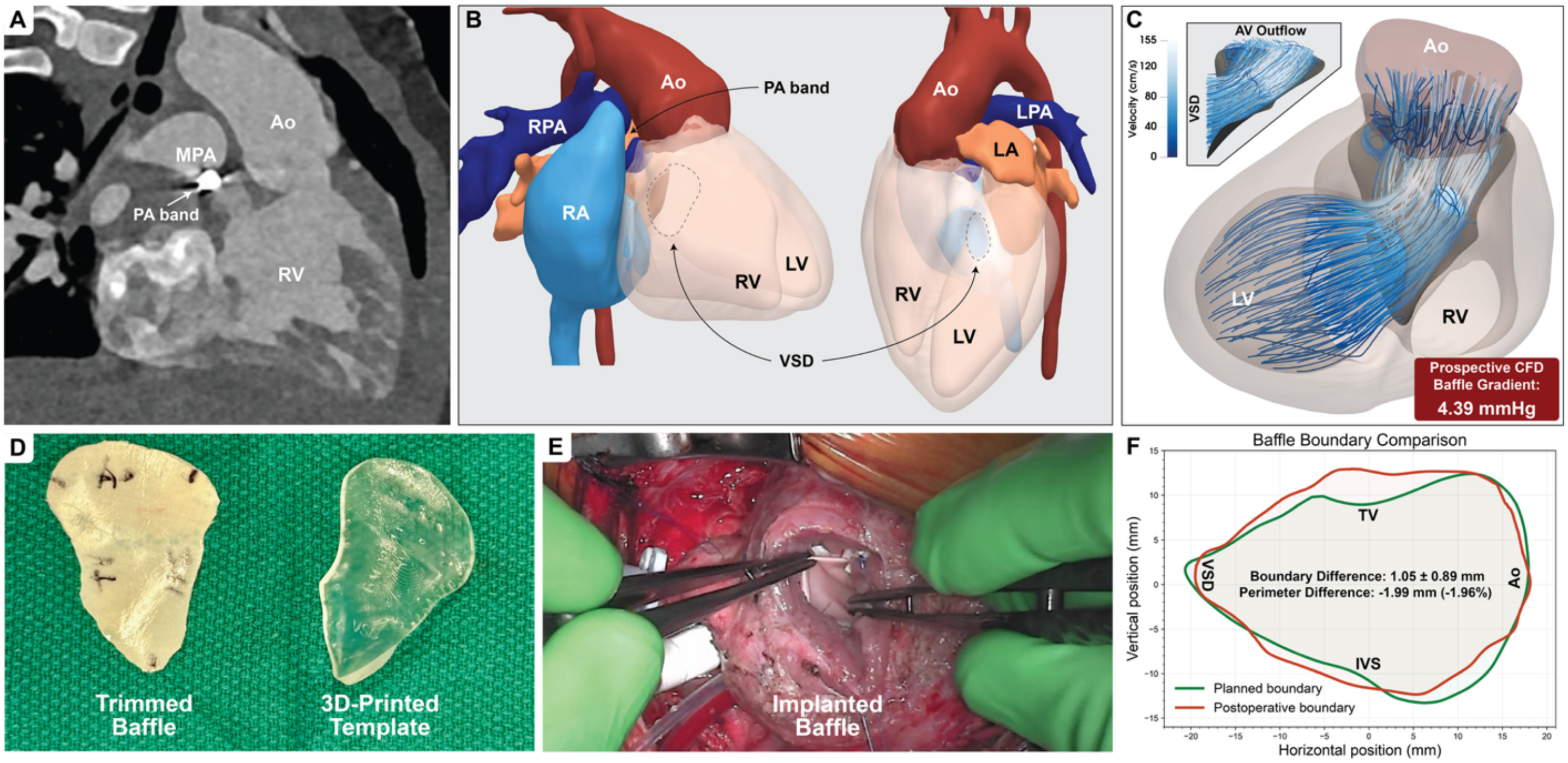

**Figure 1. Preoperative semi-automated baffle patch planning facilitates intraoperative baffle construction and generates physiologically appropriate gradients.** A. Multiplanar reconstruction of preoperative CTA demonstrating double outlet right ventricle with a subpulmonic VSD and pulmonary artery band. B. Three-dimensional computational model from patient's preoperative CTA demonstrating a reasonable left ventricle to aorta pathway. C. Preoperative computational fluid dynamic analysis of baffle template with velocity streamlines mapped over neo-left ventricular outflow tract pathway. Inset represents an orthogonal view of the neo-left ventricular outflow tract. D. Trimmed and labeled bovine pericardial patch adjacent to 3D-printed baffle template prior to implantation. E. Implanted computationally guided baffle viewed through right ventriculotomy. F. Overlaid preoperative template (green) and implanted baffle (red) projected onto a shared, least-squares best-fit plane following alignment of the VSD and aortic outlets from the pre- and postoperative geometries. VSD: ventricular septal defect;

CTA: computed tomography angiogram; 3D: three-dimensional; TV: tricuspid valve; Ao: aortic outflow; IVS: interventricular septum.

**References**

1. Martinez ES, Kaiser AD, Reed AK, et al. Semi-Automated Generation and Hemodynamic Assessment of Surgical Baffle Geometry for Biventricular Repair. *arXiv*. Preprint posted online September 10, 2026. doi:10.48550/ARXIV.2603.25207

2. Cattermole GN, Leung PYM, Mak PSK, Chan SSW, Graham CA, Rainer TH. The normal ranges of cardiovascular parameters in children measured using the Ultrasonic Cardiac Output Monitor: *Crit Care Med*. 2010;38(9):1875-1881. doi:10.1097/CCM.0b013e3181e8adee

3. Hoganson DM, Govindarajan V, Schulz NE, et al. Multiphysiologic State Computational Fluid Dynamics Modeling for Planning Fontan With Interrupted Inferior Vena Cava. *JACC Adv*. 2024;3(7):101057. doi:10.1016/j.jacadv.2024.101057

4. Liang J, Fang X, Li D, Pan G, Zhang G, Lu B. Application feasibility of virtual models and computational fluid dynamics for the planning and evaluation of aortic repair surgery for Williams syndrome. *J Cardiothorac Surg*. 2025;20(1):144. doi:10.1186/s13019-024-03286-6

5. Vigil C, Lasso A, Ghosh RM, et al. Modeling Tool for Rapid Virtual Planning of the Intracardiac Baffle in Double-Outlet Right Ventricle. *Ann Thorac Surg*. 2021;111(6):2078-2083. doi:10.1016/j.athoracsur.2021.02.058

6. Hasegawa T, Yamaguchi M, Yoshimura N, Okita Y. The dependence of myocardial damage on age and ischemic time in pediatric cardiac surgery. *J Thorac Cardiovasc Surg*. 2005;129(1):192-198. doi:10.1016/j.jtcvs.2004.05.005